\documentclass[a4paper,12pt]{article}
\pdfoutput=1
\usepackage{jcappub}
\usepackage{amsmath,amssymb,amsfonts,color,epsf,epsfig,epstopdf,graphics,graphicx,mathtools,subcaption,physics,verbatim}
\newcommand{\nc}{\newcommand}
\nc{\ba}{\begin{eqnarray}}
\nc{\ea}{\end{eqnarray}}

\newcommand{\calP}{{\cal{P}}}
\newcommand{\p}{{\partial}}
\newcommand{\Mpl}{M_{\rm Pl}}
\newcommand{\A}{A^T}
\newcommand{\m}{m_{\tiny A}}
\newcommand{\f}{f}
\newcommand{\h}{h}
\newcommand{\vp}{\varphi}

\newcommand{\eq}[1]{\begin{equation}#1\end{equation}}
\newcommand{\eqa}[1]{\begin{align}#1\end{align}}

\newcommand{\fg}[1]{\begin{figure}[tbp]\centering #1 \end{figure}}

\nc{\e}{{\bf{e}}}

\title{  {Dark photon dark matter from charged inflaton}}

\author[a]{Hassan Firouzjahi,}
\author[b]{Mohammad Ali Gorji,}
\author[b,c]{Shinji Mukohyama,} 
 \author[a]{\hspace{2 cm} Borna Salehian }

\affiliation[a]{School of Astronomy, 
	Institute for Research in Fundamental Sciences (IPM) \\ 
	P.~O.~Box 19395-5531, Tehran, Iran}
\affiliation[b]{Center for Gravitational Physics,
	Yukawa Institute for Theoretical Physics \\
	Kyoto University, 606-8502, Kyoto, Japan}
\affiliation[c]{Kavli Institute for the Physics and Mathematics of the Universe (WPI), 
	The University of Tokyo Institutes for Advanced Study, 
	The University of Tokyo, Kashiwa, Chiba 277-8583, Japan}

\emailAdd{firouz@ipm.ir}
\emailAdd{gorji@yukawa.kyoto-u.ac.jp}
\emailAdd{shinji.mukohyama@yukawa.kyoto-u.ac.jp}
\emailAdd{salehian@ipm.ir}

\abstract{We present a scenario of vector dark matter production during inflation  containing a complex inflaton field  which is charged under a dark gauge field and which has a symmetry breaking potential.  As the inflaton field rolls towards the global minimum of the potential the dark photons become massive with a mass which can be larger than the Hubble scale during inflation. The accumulated energy of the quantum fluctuations of the produced dark photons gives the observed relic density of the dark matter for a wide range of parameters. Depending on the parameters, either the transverse modes or the longitudinal mode or their combination  can generate the observed dark matter relic energy density. }

\begin{document}
\begin{flushright} {\footnotesize YITP-20-139, IPMU-0115}  \end{flushright}
\maketitle
\flushbottom

\section{Introduction} \label{sec:intro}
There are numerous compelling evidences for the existence of dark matter but understanding its nature remains one of the important questions in cosmology. 
Dark photon dark matter, alternatively known as ``vector dark matter'', is one of the candidate for dark matter which is  widely studied in recent years \cite{Nelson:2011sf,Arias:2012az,Graham:2015rva,Agrawal:2017eqm, Co:2018lka, Agrawal:2018vin,Dror:2018pdh}. It can be produced before the time of matter and radiation equality through their direct interactions with other particles in the dark sector or with the inflaton field 
and then playing the role of dark matter after the time of matter and radiation equality. It should have negligible direct interactions with the Standard Model (SM) particles in order not  to destroy the predictions of the SM \cite{Goodsell:2009xc,Alexander:2016aln}. However, dark photon dark matter can be indirectly probed by means of the gravitational waves observations through its inevitable minimal coupling with gravity \cite{Pierce:2018xmy,Machado:2018nqk,Machado:2019xuc,Salehian:2020dsf,Michimura:2020vxn,Namba:2020kij,Kitajima:2020rpm}. 

The vector dark matter models are usually constructed with a mechanism for the production of dark photons after inflation but before the time of  matter and radiation equality \cite{Agrawal:2017eqm,Co:2017mop,Agrawal:2018vin,Dror:2018pdh}. However,  it is noticed that they can be produced even during inflation from the longitudinal mode of a massive gauge boson minimally coupled to the inflaton field \cite{Graham:2015rva, Kolb:2020fwh}. For the transverse modes, while their energy density usually decays exponentially, such a decay can be avoided and they can instead be produced if the gauge field is coupled non-minimally to the inflaton field \cite{Anber:2006xt,Watanabe:2009ct}. In this way, the dark gauge field acquires energy from the inflaton field and the transverse modes can be produced efficiently during inflation to play the roles of dark matter after the time of matter and radiation equality \cite{Bastero-Gil:2018uel,Nakayama:2019rhg,Nakayama:2020rka,Nakai:2020cfw}. For this purpose, we need a large enough mass for dark photons so that they become non-relativistic before the time of matter and radiation equality. On the other hand, having a large mass compared to the Hubble scale causes an inefficient particle production during inflation. That is why most of the previous studies assumed a very small mass for dark photons. In Ref. \cite{Salehian:2020asa} we have circumvented these issues by generating a dynamical mass for the dark photon at the end of inflation through the symmetry breaking  i.e. the Higgs mechanism. In \cite{Salehian:2020asa}, a real scalar field is the inflaton field while a massive complex field plays the roles of symmetry breaking and terminating inflation, much like the waterfall field of hybrid inflation. As the heavy complex field  rolls rapidly to the global minimum of its potential a mass is generated for the vector field via the Higgs mechanism. The induced mass can be large such that the excitations of the transverse modes can furnish the current dark matter relic energy density. An important effect in \cite{Salehian:2020asa} was that the longitudinal mode can not contribute to the dark matter relic density as it is non-dynamical during inflation while being too heavy to be produced efficiently at the end of inflation.  

In this work we consider a model of inflation in which the inflaton itself is charged under the dark photon. The potential near the minimum is similar to a simple chaotic model but with a non-zero vacuum expectation value (vev). In order to amplify the dark photon excitations, as in \cite{Salehian:2020asa}, we allow a coupling between the radial component of the inflaton field and the dark photon through the gauge kinetic function. As the inflaton rolls towards its minimum  with a non-zero initial value, the dark photons acquire mass due to the Higgs mechanism. Consequently, unlike \cite{Salehian:2020asa}, the longitudinal mode becomes dynamical during inflation and can contribute to the dark matter relic density.  In a sense, this setup may be viewed as a hybrid of \cite{Salehian:2020asa} where dark matter relic comes from  the energy density of transverse modes and \cite{Graham:2015rva} where only the longitudinal mode is the origin of dark matter. 

The rest of the paper is organized as follows. In section \ref{sec-SB} we present our setup and in section \ref{sec-Model}  we specify the form of the gauge kinetic function and solve the mode functions. In section  \ref{relic-sec} we calculate the relic energy density followed by a summary and discussions in section  \ref{summary}. Some technicalities associated with the longitudinal mode are relegated to the Appendix.

\section{Inflationary model with Higgs mechanism}\label{sec-SB}

We consider an inflationary model in which the inflaton is a complex scalar field $\phi$ charged under a $U(1)$ gauge field  $A_\mu$. We assume no coupling between the gauge field and the SM particles, namely it is a dark photon. The theory is described by the following action
\begin{eqnarray}
\label{action} S = \int
d^4 x  \sqrt{-g} \bigg[ \frac{\Mpl^2}{2} R
- \frac{1}{2} D_\mu\phi \overline{D^\mu\phi} - V(|\phi|) 
- \frac{1}{4} \f^{2}(|\phi|) F_{\mu \nu} F^{\mu \nu} \bigg] \,,
\end{eqnarray}
where $\Mpl = 1/ \sqrt{8 \pi G}$ is the reduced Planck mass and $G$ is the Newton constant. The components of the covariant derivative acting on the inflaton field are given by
\ba
D_\mu
\phi = \p_\mu  \phi + i \e \,  \phi  \, A_\mu \,,
\ea
in which $\e$ is the charge of the inflaton field which is a dimensionless quantity. The components of the field strength tensor associated to the gauge field is given by $F_{\mu\nu}=\p_\mu A_\nu - \p_\nu A_\mu$. Furthermore, we have introduced the gauge kinetic function $f(|\phi|)$ so that the energy can be transferred from the inflaton to the gauge field. This form of coupling is extensively used in the context of anisotropic inflation \cite{Watanabe:2009ct,Watanabe:2010fh,Emami:2010rm,Bartolo:2012sd, Emami:2015qjl,Shakeri:2019mnt} and also primordial magnetogenesis \cite{Ratra:1991bn,Garretson:1992vt,Anber:2006xt,Martin:2007ue,Demozzi:2009fu,Kanno:2009ei,Emami:2009vd,Fujita:2012rb,Caprini:2014mja,Caprini:2017vnn,Schober:2020ogz,Talebian:2020drj}.

By assuming the $U(1)$ gauge symmetry, the potential and the gauge kinetic function depends on only the modulus of the inflaton field $|\phi|$. We decompose the inflaton into the radial and angular components as follows
\ba
\label{phi}
\phi \equiv \vp e^{i \theta}\,,
\ea
where $\theta$ determines the argument of the complex inflaton field $\phi$ while $\varphi$ characterizes the amplitude $\varphi=|\phi|$. By using the local gauge redundancy of the theory we choose $\theta=0$, i.e. the unitary gauge, and deal with a real and positive valued inflaton field $\vp$. The gauge kinetic function and the potential are then only a function of $\vp$. We will not assume any specific form for $f(\vp)$ in this section. However, we assume a power law form for it as a function of conformal time in the next sections.   

For the potential we do not assume any specific functional form other than a couple of general features. First, the potential must satisfy the slow-roll conditions to provide enough number of e-folds of accelerated expansion. Second, we assume that the  inflaton field settles finally into its global minimum with $V=0$ with a  a nonzero vev. This is to generate a mass for the dark photon, as in  Higgs mechanism, which is related to the inflaton field as \cite{Salehian:2020asa}
\ba
\label{m-DF}
\m^2= \e^2 \vp^2\,.
\ea    
Since the dark photon is meant to form the observed dark matter content of the universe, it has to be massive after inflation. As a result, we need to have $\vp\neq0$ at the end of inflation. For later usage, we define the ratio of the gauge field mass to the Hubble scale during inflation as 
\eq{
	\label{R}
	R\equiv\frac{\m^2}{H^2}=\frac{\e^2\vp^2}{H^2}\,.
}
Note that one can tune parameters of the theory such as $\e$ so that  dark photons can be heavier ($R>1$) or lighter ($R<1$) than the Hubble scale during inflation. Further, the parameter $R$ generically depends on time. More specifically  we have
\eq{
\label{Rdot}
\frac{\dot{R}}{HR}=\left(\sqrt{2\epsilon}+\frac{2\Mpl}{\vp}\right)\sqrt{2\epsilon}\,,
} 
where $\epsilon \equiv -\dot{H}/H^2$ is the slow roll parameter and a dot denotes the derivative with respect to the cosmic time. If we assume $\vp\sim\order{\Mpl}$, which is in fact the case in many inflationary models, then Eq.~\eqref{Rdot} shows that the fractional change of the parameter $R$ in one e-fold is slow-roll suppressed. As we neglect slow-roll corrections in the following analysis, we assume that $R$ does not change with time. Including the effects of the running of the parameter $R$ will add subleading corrections to our results.

One can easily build a potential with the above mentioned features.  In the following subsection we consider a simple example of such inflationary model. 

\subsection{Inflaton dynamics}\label{hilltop}

As a simple explicit realization of the above setup, we consider the standard symmetry breaking potential of the form
\begin{equation}
\label{pot}
V(\vp) = \frac{\lambda}{4} \left(\vp^2 - v^2 \right)^2 \, ,
\end{equation}
in which $\lambda$ is a dimensionless coupling and $v$ is the position of the global minimum of the potential. The above potential is  well motivated in the context of symmetry breaking when a complex scalar field is charged under a $U(1)$ field, such as in the Higgs symmetry breaking mechanism. The potential is depicted in Fig.~\ref{fig:higgs}. We can imagine two types of inflationary scenarios for $\vp<v$ and $\vp>v$. We will review these two types in the following. Here, we do not consider them as realistic inflationary models as their predictions are disfavoured by the Planck observation \cite{Akrami:2018odb}.  Instead, since the computations in the next sections do not depend on the details of inflationary models, we use these simple models just to show that generically $R$ can be large. 
\fg{
	\centering
	\includegraphics[width=0.33\textwidth]{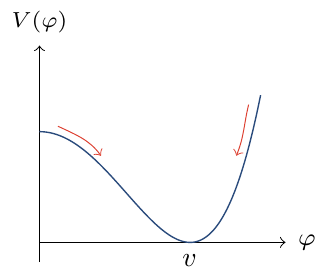}
	\caption{A schematic view of the potential $V(\vp)$. One can imagine two inflationary scenarios depending on $\vp<v$ or $\vp>v$.}
	\label{fig:higgs}
}

\vspace{3mm}

{\bf Hilltop case.} First consider the case where inflation proceeds as in hilltop model in the region  $0< \vp < v$. In this region, the background expansion is given by 
\ba
3 \Mpl^2 H^2 \simeq \frac{\lambda v^4}{4},
\ea
and the number of e-folds needed to solve the flatness and the horizon problems is given by
\ba
\label{N-1}
\mathcal{N}= \frac{1}{\Mpl^2} \int d \varphi \, \frac{ V}{V_\varphi} \simeq 
\frac{v^2}{4 \Mpl^2} \ln{\frac{\vp_{\rm e}}{\vp_{\rm i}}}\, ,
\ea
where $\vp_{\rm i}$ and $\vp_{\rm e}$ are the initial and the final values of the 
inflaton field respectively. 
We expect that $\vp_{\rm e}$ is not exponentially larger than $\vp_{\rm i}$, i.e. $\vp_{\rm e}\sim\vp_{\rm i}$ up to an $\order{1}$ factor, so one typically has $\mathcal{N} \sim v^2/4 \Mpl^2$.  

The slow-roll conditions require that $v>2\Mpl$ and $\vp<v/\sqrt{2}$ for a successful inflation model. One can compute the change in the parameter $R$ during inflation as
\eq{
\label{ReRi1}
\frac{R_{\rm e}}{R_{\rm i}}=\left(\frac{H_{\rm i}}{H_{\rm e}}\right)^2
\left(\frac{\vp_{\rm e}}{\vp_{\rm i}}\right)^2\simeq\left(\frac{\vp_{\rm e}}{\vp_{\rm i}}\right)^2\,,
}  
which as discussed above is an $\order{1}$ value. Finally, the parameter $R$ at the end of inflation can be expressed as
\ba
\label{R-val}
R_{\rm e} = \frac{\e^2 \vp_{\rm e}^2}{H_{\rm e}^2} \simeq \frac{\e^2}{4\pi^2\calP}\left(\frac{v}{2\Mpl}\right)^4\sim\frac{\e^2\mathcal{N}^2}{4\pi^2\calP}\,,
\ea
where we have used the definition of the dimensionless curvature perturbation 
power spectrum as
\ba
\label{power0}
\calP = \frac{H^2}{8 \pi^2 \epsilon \Mpl^2}\simeq 2\times 10^{-9}\,.
\ea 
The relation \eqref{R-val} shows that with $\e \sim 1$, the parameter $R$ can be as large as $10^{12}$. By reducing $\e$, smaller values of $R$ can be obtained. For example, with $\e \sim 10^{-3}$ we obtain $R \sim 10^6$. 

\vspace{3mm}

{\bf Large field.} Now let us consider the case of large field model where $\vp > v$ so the field rolls from the large value of $\vp$ toward its global minimum at $\vp=v$. For large values of $\varphi$, the potential looks like $V \simeq  \lambda  \varphi^4/4$ so inflation proceeds as in chaotic models with large field excursions. Of course, these types of large field models are disfavoured by the Planck data as they produce large tensor perturbations \cite{Akrami:2018odb}. Here, however, we presented this setup as a simple working example for the purpose of dark photon and symmetry breaking while a variant of this potential can be invoked to be consistent with the Planck constraint on primordial gravitational waves as well.

The number of e-fold to solve the flatness and the horizon problems in this case is given by
\ba
\label{N-11}
\mathcal{N}= \frac{1}{\Mpl^2} \int d \varphi \, \frac{V}{V_\varphi} \simeq 
\frac{1}{8 \Mpl^2}(\varphi_{\rm i}^2 - \varphi_{\rm e}^2) \, .
\ea
Note that $\varphi_{\rm i} > \varphi_{\rm e} >  v$. Assuming $\vp_{\rm i}\gg\vp_{\rm e}$ we approximately have $\mathcal{N}\sim\vp_{\rm i}^2/8\Mpl^2$ and the change in the parameter $R$ is easily computed to be
\eq{
\label{ReRi2}
\frac{R_{\rm e}}{R_{\rm i}}\simeq\left(\frac{\vp_{\rm i}}{\vp_{\rm e}}\right)^2\sim\mathcal{N}\,,
} 
which shows that $R$ changes as the logarithm of the conformal time. As we will see, this running can be neglected compared to other terms involved in the model in agreement with the general argument \eqref{Rdot}. Also, the value of $R$ at the start of inflation is obtained to be 
\ba
R_{\rm i} = \frac{\e^2 \varphi_{\rm i}^2}{H_{\rm i}^2} = \frac{\e^2}{8 \pi^2 \calP \epsilon} 
\frac{\varphi_{\rm i}^2}{\Mpl^2} \simeq \frac{\e^2 \mathcal{N}^2}{\pi^2 \calP} \,.
\ea
Thus, as in the case of hilltop inflation,  $R_{\rm i}$ can be very large depending on the value of $\e$. 

\vspace{3mm}

We comment that the standard symmetry breaking potential Eq. (\ref{pot}) may be used as a working example for our inflationary dynamics. However, as we mentioned before, other potentials can be used too as long as the potential has a non-zero vev at the end of inflation to generate mass for the dark photon. In this way, we can avoid the strong bounds from the Planck observation on potential  (\ref{pot}) such as the bound on the tensor-to-scalar power spectra ratio.  As we shall see in next sections  the details of the relic density will not be strongly dependent on the shape of the potential at the start of inflation.  
Furthermore, as discussed above, we neglect the time dependence of $R$ in the following analysis so hereafter we write $R$ without any reference to the corresponding time (with the index ${\rm i}$ or ${\rm e}$).

Contrary to the model discussed in \cite{Salehian:2020asa}, where the mass of the vector field is generated by the vev of the waterfall field after hybrid inflation, in the present model the $U(1)$ symmetry breaking does not result in the formation of cosmic strings. This is because, in the present model, a single value of the phase $\theta$ (up to quantum fluctuations) is chosen by the inflationary trajectory in the observable patch of the universe. As a result, we do not have to impose the CMB constraint on the tension of cosmic strings.

In the following subsection we focus on the gauge field sector of the theory to study the process of dark photon production.   

\subsection{Vector field dynamics}\label{sec-DP}

As we explained in the previous section, due to the Higgs mechanism, the dark gauge field acquires mass during inflation which depends on the inflaton field as given in Eq. (\ref{m-DF}). Therefore, the dark photon sector of the action \eqref{action} would take the following form
\ba
\label{actionA} S_{\tiny A}= \int
d^4 x  \sqrt{-g} \left [ - \frac{1}{4} \f^{2}(\vp) F_{\mu\nu} F^{\mu\nu}
- \frac{1}{2}\m^2 \, A_\mu A^\mu \right]\,.
\ea
The inflationary background is given by the spatially flat Friedmann-Lema\^{i}tre-Robertson-Walker (FLRW) metric as
\begin{align}\label{FLRW}
{\rm d}s^2 = a^2(\tau) \big(-{\rm d}\tau^2+\delta_{ij}~{\rm d}x^i{\rm d}x^j \big) \, ,
\end{align}
where $a(\tau)$ is the scale factor and $\tau$ is the conformal time. Taking the advantage of the rotational symmetry of the spatial part of the metric, we decompose the spatial components of the gauge field as
\ba \label{dec-A}
A_i=\p_i\chi+\A_i\,,\hspace{1cm}\delta^{ij}\p_i\A_j=0\,,
\ea
where $\chi$ and $\A_i$ characterize the longitudinal and transverse parts of the gauge field respectively. The component $A_0$ is non-dynamical and it can be solved in terms of the longitudinal mode as \cite{Salehian:2020asa}
\eq{
	\label{A0}
A_0=\Big(\frac{\f^2\nabla^2}{\f^2\nabla^2-\m^2a^2}\Big)\chi'\,,
}
where a prime denotes the derivative with respect to the conformal time. 

Substituting the above solution in the action, it can then be decomposed into the transverse and longitudinal parts as
\eq{
	S_{\tiny A} = S_T+S_L\,,
}
where
\eqa{\label{ST}
	S_T&=\frac{1}{2}\int\dd[3]{x}\dd{\tau}\Big[ \f^2 (\A_i{}')^2
	- \f^2 (\p_i\A_j)^2-\m^2a^2(\A_i)^2 \Big] \,, \\
	\label{SL}
	S_L&=\frac{1}{2}\int\dd[3]{x}\dd{\tau}\m^2a^2
	\Big[\chi'\Big(\frac{\f^2\nabla^2}{\f^2\nabla^2-\m^2a^2}\Big)\chi'-(\p_i\chi)^2\Big] \,.
}
Finally, the total energy density of the gauge field during inflation can be decomposed as $\rho_{\tiny A}=\rho_T+\rho_L$ where $\rho_T$ and $\rho_L$ are the energy densities for the transverse and longitudinal modes respectively given by (see Ref. \cite{Nakai:2020cfw,Salehian:2020asa} for the details)
\eq{
\label{rhoT}
\rho_T=\frac{\f^2}{2a^4}\Big[(\A_i{}')^2+(\p_i\A_j)^2+\frac{\m^2a^2}{\f^2}(\A_i)^2\Big]\,,\quad
\rho_L=\frac{\m^2}{2a^2}\Big[\chi'\big(\frac{\f^2\nabla^2}{\f^2\nabla^2-\m^2a^2}\big)\chi'+(\p_i\chi)^2\Big]\,.
} 
The energy density of the vector field is the accumulated energy of its quantum fluctuations for both transverse and longitudinal modes which are produced during inflation. 

The quantization procedure follows almost exactly the standard textbooks. For the transverse modes we define the canonical field $V_i\equiv f\A_i$ and expand it in terms of the creation  and annihilation operators $a^\dagger_{\vb{k},\lambda}$ and $a_{\vb{k},\lambda}$ as usual
\begin{equation}\label{V-mode}
V_i=\sum_\lambda\int\frac{\dd[3]{k}}{(2\pi)^{3/2}} \varepsilon^\lambda_i({\bf k})
\big[v_{k}(\tau)a_{\vb{k},{\lambda}}
+v_{k}(\tau)^*a^\dagger_{-\vb{k},{\lambda}} \big] e^{i\vb{k}.\vb{x}}\,, \hspace{1cm} 
[a_{\vb{k},{\lambda}}, a^\dagger_{\vb{k}',{\lambda'}}] = \delta_{\lambda\lambda'}\delta(\vb{k}-\vb{k}')\,,
\end{equation}
where the time dependence of the vector field is encoded in the mode function $v_{k}(\tau)$. In the above relation, $\varepsilon^\lambda_i({\bf k})$ are the polarization vectors for $\lambda =\pm$ (see appendix A of Ref. \cite{Salehian:2020dsf} for properties of polarization vectors). The mode function satisfies the equation of motion 
\ba\label{MF-Eq}
v_{k}'' + \Big( k^2-\frac{\f''}{\f}+ \frac{\m^2a^2}{\f^2}\Big) v_{k} =0   \, ,
\ea 
subject to the Bunch-Davies initial condition.

Similarly, for the longitudinal mode we define the canonical field
\ba
\label{U}
U\equiv h\, \chi\,,\qquad h \equiv \m a f\sqrt{\frac{\nabla^2}{f^2\nabla^2-\m^2a^2}}\,.
\ea
Then we expand the field $U$ in terms of the creation and annihilation operators $b^\dagger_{\vb{k}}$ and $b_{\vb{k}}$ as follows 
\begin{equation}\label{U-mode}
U=\int\frac{\dd[3]{k}}{(2\pi)^{3/2}}
\big[u_{k}(\tau)b_{\vb{k}}
+u_{k}(\tau)^*b^\dagger_{-\vb{k}} \big] e^{i\vb{k}.\vb{x}}\,, \hspace{1cm} 
[ b_{\vb k} , b_{\vb{k}'}^\dagger] =\delta(\vb{k}-\vb{k}')\,,
\end{equation}
and the equation for the mode function is easily found to be
\ba\label{MF-EqL}
u_{k}'' + \Big( k^2-\frac{h''}{h}+ \frac{\m^2a^2}{\f^2}\Big) u_{k} =0   \, ,
\ea 
which is also subject to the Bunch-Davies initial condition. 

Finally, the energy density of the transverse modes and the longitudinal mode can be computed from Eq.~\eqref{rhoT} in terms of the corresponding mode functions
\eq{
	\label{rhoT3}
	\rho_T=\frac{1}{a^4}\int\frac{\dd[3]{k}}{(2\pi)^3}
	\Big[ \big|{ \f\big(\frac{v_{k}}{\f}\big)'} \big|^2+\big(k^2+\frac{\m^2a^2}{\f^2}\big)\abs{v_{k}}^2\Big]\,,
}  
\eq{
	\label{rhoT4}
	\rho_L=\frac{1}{2a^4}\int\frac{\dd[3]{k}}{(2\pi)^3}
	\Big[ \big|{ h\big(\frac{u_{k}}{h}\big)'} \big|^2+\big(k^2+\frac{\m^2a^2}{\f^2}\big)\abs{u_{k}}^2\Big]\,.
}  
We will be interested in the total energy density at the end of inflation so it is  useful to parameterize the energy density at that time as follows
\eq{
\label{rhoC}
\rho_T(\tau_{\rm e})=\mathcal{C}_T\, H_{\rm e}^4\,,\qquad\rho_L(\tau_{\rm e})=\mathcal{C}_L\, H_{\rm e}^4\,,
}
where $H_{\rm e}$ is the Hubble expansion rate at the end of inflation and $\mathcal{C}_T$ and $\mathcal{C}_L$ are dimensionless functions which depend on the parameters of our model. The energy density of the produced dark photons at the end of inflation is $\rho_{\rm e}=\rho_T(\tau_{\rm e})+\rho_L(\tau_{\rm e})$ and we similarly define $\rho_{\rm e} \equiv \mathcal{C}\, H_{\rm e}^4$ 
with $\mathcal{C} \equiv \mathcal{C}_T+\mathcal{C}_L$. 

So far in our analysis, we did not specify the form of the gauge kinetic function $\f$. In the following section, we assume a specific form for the gauge kinetic function and derive an explicit expression for $\mathcal{C}_T$ and $\mathcal{C}_L$. Then we show that such a model can account for the observed dark matter energy density.

\section{The phenomenological model}\label{sec-Model}

As we mentioned before, during inflation the gauge field energy density usually decays quickly. This issue can be remedied if energy is pumped from the inflaton field to the gauge field through the coupling $\f(\vp)$. Since in single clock inflation the inflaton field can be exchanged by time, the coupling $\f$ can be written equivalently as a function of time. To simplify the analysis we assume that the time-dependence is of the form of a power law as
\begin{eqnarray}\label{f1}
\f(\tau) =\left(\dfrac{\tau}{\tau_{\rm e}}\right)^{p}\,,
\end{eqnarray}
where we have normalized the coupling so that $\f(\tau_{\rm e})=1$. In order to have a well-defined perturbative theory, we require $\f^{-1} <1$ during inflation which implies $p>0$. To simplify the analysis further, we assume that after inflation the dependence of $\f$ on $\vp$ is sufficiently weak such that we can safely ignore any particle production during (p)reheating. Note that the gauge kinetic function in the form of Eq. (\ref{f1}) has been employed extensively in the context of magnetogenesis \cite{Ratra:1991bn,Garretson:1992vt,Anber:2006xt, Martin:2007ue, Demozzi:2009fu,Kanno:2009ei,Emami:2009vd,Fujita:2012rb,Caprini:2014mja,Caprini:2017vnn, Schober:2020ogz,Talebian:2020drj}, anisotropic inflation  \cite{Watanabe:2009ct, Watanabe:2010fh, Bartolo:2012sd, Emami:2010rm, Emami:2013bk,Abolhasani:2013zya,Emami:2015qjl,Shakeri:2019mnt}, pseudoscalar inflaton \cite{Anber:2006xt,Sorbo:2011rz}, and inflation with multiple vector fields \cite{Yamamoto:2012tq,Yamamoto:2012sq,Firouzjahi:2018wlp,Gorji:2020vnh}.

In the following we  solve the mode functions for the transverse and longitudinal modes which will be used to calculate their contributions to the relic density of dark matter. 

\subsection{Transverse modes}

Assuming the specific form (\ref{f1}) for the gauge kinetic function, the equation for the transverse mode function Eq.~\eqref{MF-Eq} takes the following form
\ba
\label{v-eq1}
v_{k}'' + \Big( k^2 -\frac{p (p-1)}{\tau^2} +\frac{R}{\tau^2}\left(\frac{\tau_{\rm e}}{\tau}\right)^{2p}\Big) v_{k}=0\,. 
\ea
As we have discussed previously the time dependence of the parameter $R$ can be ignored up to slow roll corrections and it can be treated as a constant throughout the period of inflation. Note that the value of $R$ can be much larger than unity.

From Eq.~\eqref{v-eq1} we see the competition between the last two terms in the big bracket. The term containing $p(p-1)/\tau^2$ is from the gauge kinetic function and for the particular case $p=2$ yields the usual term $2/\tau^2$ in the equation of motion for the mode function of a massless scalar field which is the hallmark of the scale invariant perturbations in a fixed de Sitter background.  If $R$ is chosen large enough, then the last term can dominate over the second term towards the final stage of inflation. This happens at the time $\tau_{\rm c}$ when the two terms become equal \cite{Emami:2009vd}
\eq{
\label{etac}
\frac{\tau_{\rm c}}{\tau_{\rm e}}=\Big(\frac{R}{p(p-1)}\Big)^{\frac{1}{2p}}\,.
}   

If $R$ is large enough then the mass term of the dark photon plays important roles in the evolution of mode functions. Specifically, assuming $R>p(p-1)$ hereafter, the dominance of the mass term happens sometime during inflation. To have a sense as when the effect of mass becomes important it is better to express Eq. (\ref{etac}) in terms of the number of e-folds. Defining the number of e-folds from the time when the condition (\ref{etac}) is met
 till the time of end of inflation  by $\Delta \mathcal{N}$, we obtain
\ba
\label{DeltaN}
\Delta\mathcal{N} \equiv \ln\frac{\tau_{\rm c}}{\tau_{\rm e}} 
= \frac{1}{2p} \ln \Big(  \frac{R}{p (p-1)}\Big) \, .
\ea
The larger $\Delta\mathcal{N}$ is, the sooner the effect of mass becomes important. For example for $p=2$ and $R= 10^{12}$ we obtain $\Delta\mathcal{N} \simeq 7$, so only towards the last seven e-folds the mass term becomes important. Of course, with smaller values of $R$, $\Delta\mathcal{N}$ becomes smaller. 

Technically speaking, the effect of the mass in the mode function is captured by the combination\footnote{The vector field defined in the action \eqref{actionA} is not canonically normalized. After defining the canonical field all couplings receive a factor of $f^{-1}$.} $m_A^2 a^2/f^2$ as is evident from Eq. (\ref{MF-Eq}). Since $f$ is very large initially and become $\order{1}$ only at the end of inflation, the effects of the mass of dark photon are subleading during inflation so they can be excited efficiently during most of the period of inflation.   Thus, even for dark photons much heavier than the Hubble scale with $R \gg 1$,  we can have efficient particle production during inflation. Finally, note that $\tau_{\rm c}$ and $\Delta\mathcal{N}$ are independent of scale. Hence, the domination of the mass term happens at the same time for all momenta. 

From the above discussion, one can ignore the last term in the brackets of Eq.~\eqref{v-eq1} at early times during inflation. Starting with the  Bunch-Davies initial condition, the solution during this stage  is given by
\ba
\label{mode1}
v_k (\tau) = \frac{\sqrt{-\pi \tau}}{2} e^{i (1+ 2 \nu)\pi/4} H_\nu^{(1)} (x) \, ,\qquad\text{early times, } -\tau\gg-\tau_{\rm c}\,,
\ea  
in which $\nu \equiv p-1/2$, $H_\nu^{(1)}$ is the Hankel function of the first kind and its argument is $x\equiv-k\tau$. On the other hand, towards the final stage of inflation  the last term in Eq.~\eqref{v-eq1} containing $R$ dominates and the solution is given by \cite{Emami:2009vd}
\ba 
\label{mode2} v_k(\tau) =
\frac{\sqrt{-\pi  \tau}}{2}e^{i ( 1+ 2 \nu ) \pi/4} \left[ c_2 \, H_{\mu}^{(1)}(
y)+ c_1 \, H_{\mu}^{(2)}( y) \right] \, ,\qquad\text{late times, } -\tau\ll-\tau_{\rm c}\,,
\ea  
where $\mu \equiv 1/2p$ and $c_1$ and $c_2$ are two constants of integration and we have defined
\eq{
	y\equiv\frac{\sqrt{R}}{p}\left(\frac{\tau_{\rm e}}{\tau}\right)^{p}\,.
}
Matching the late time solution \eqref{mode2} with the early time solution \eqref{mode1} 
at $\tau =\tau_c$ fixes the coefficients $c_1$ and $c_2$. More specifically, demanding that both $v_k$ and $v_k'$ to be continuous at $\tau_{\rm c}$ yields 
\ba
\label{c12}
c_{\ell} = \frac{\mp i \pi}{4 p} \left[  py_{\rm c} 
H_\nu^{(1)}(x_{\rm c}) \,
H_\mu^{(\ell)}{}'(y_{\rm c}) + x_{\rm c} \,   H_\nu^{(1)}{}'(x_{\rm c}) \,  H_\mu^{(\ell)}(y_{\rm c}) \right] \, ,\qquad \ell=1,2
\ea
in which the upper (lower) sign is for $\ell=1$ ($\ell=2$) and the primes denote derivatives with respect to the corresponding arguments. Also from the  above definitions we have 
\eq{
x_{\rm c}=-k\tau_{\rm c}\,,\qquad y_{\rm c}=\frac{\sqrt{R}}{p}\Big(\frac{\tau_{\rm e}}{\tau_{\rm c}}\Big)^{p}=\sqrt{\frac{p-1}{p}}\,,
}
where in the last equality we have used Eq.~\eqref{etac}. Note that the coefficients $c_\ell$ depend on $k$ through the dependence on $x_{\rm c}$. Fig.~\ref{fig:modefunction} compares our analytic solution for the mode function with the numerical solution of Eq.~\eqref{v-eq1}. The plot shows that our analytic expression is in good agreement with the numerical result. 

\fg{
	\centering
	\includegraphics[width=0.65\textwidth]{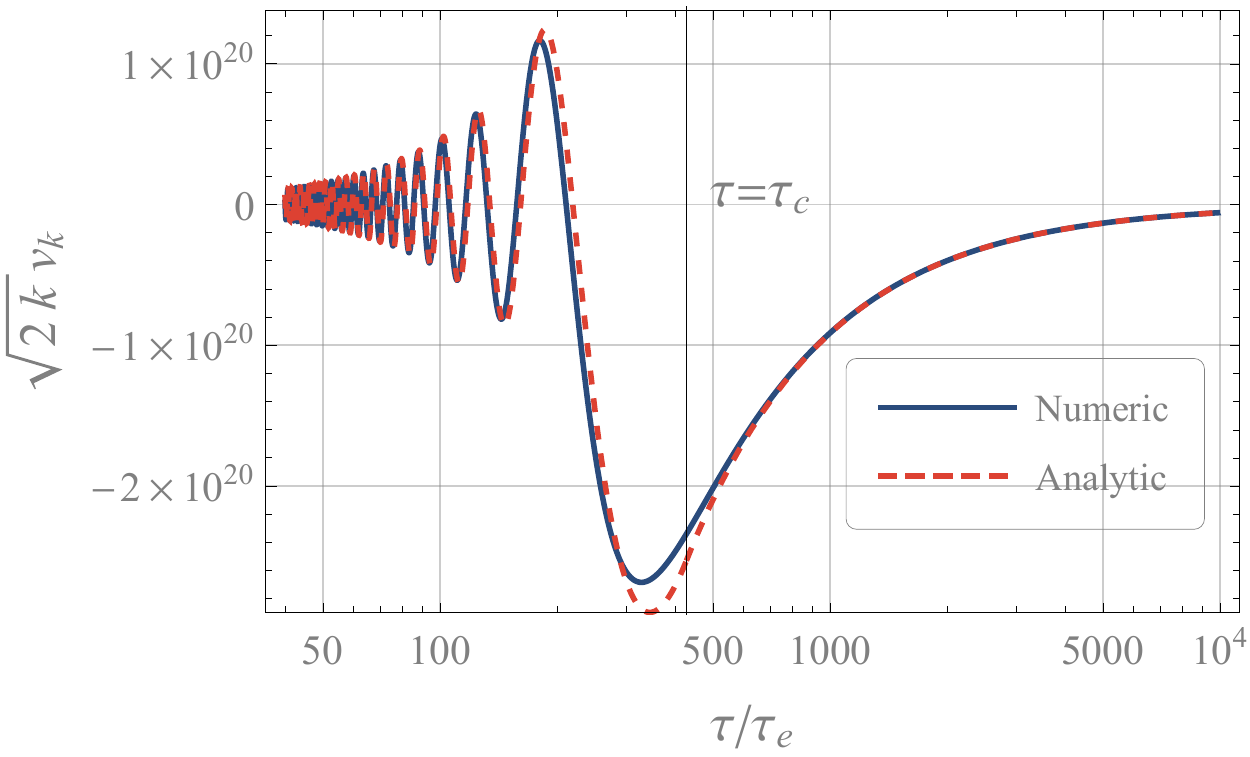}
	\caption{Comparison between the numerical solution of Eq.~\eqref{v-eq1} and the analytic expression from matching Eqs. \eqref{mode1} and \eqref{mode2} at $\tau=\tau_{\rm c}$. We have set $R=10^{12}$, $p=2.2$ and $-k\tau_{\rm e}=10^{-20}$. The latter has been chosen such that the mode lies between $k_{\rm min}$ and $k_{\rm max}$.}
	\label{fig:modefunction}
}

We are interested in the range of momenta for which dark photons can be produced via excitation of quantum fluctuations. This happens when the adiabatic condition for the mode function is broken. Most efficiently this happens if the mode function undergoes a tachyonic growth. Roughly speaking, the only chance for the mode function to become tachyonic is until the time $\tau_{\rm c}$ when the middle term in the brackets of Eq.~\eqref{v-eq1} is dominant while after that the mode cannot experience tachyonic growth. As a result, we define
\eq{
	\label{ks}
	k_{\rm min} \equiv\frac{\sqrt{p(p-1)}}{-\tau_{\rm i}}\,,\hspace{1cm} k_{\rm max}
	\equiv \frac{\sqrt{p(p-1)}}{-\tau_{\rm c}}\,,
}
corresponding to the mode that becomes tachyonic at the beginning of inflation $\tau_{\rm i}$ and at the time of $\tau_{\rm c}$ respectively. Hereafter, we assume that $\tau_{\rm i}<\tau_{\rm c}$ (i.e. $k_{\rm min}<k_{\rm max}$) so that the tachyonic growth occurs during inflation. We also assume that $R>p(p-1)$ so that $\tau_{\rm c}<\tau_{\rm e}$. Thus, dark photons with momenta in the range between $k_{\rm min}$ and $k_{\rm max}$  can be produced during inflation.    

We need to compute the energy density of the transverse modes at the end of inflation. We use Eq.~\eqref{rhoT3}, which for clarity is repeated as
\eq{
\label{energy-T}
\rho_T(\tau_{\rm e})\approx\frac{1}{a_{\rm e}^4}\int_{k_{\rm min}}^{k_{\rm max}}\frac{\dd[3]{k}}{(2\pi)^3}
	\Big[ \big|{\big(\frac{v_{k}}{\f}\big)'} \big|^2+\m^2a_{\rm e}^2\abs{v_{k}}^2\Big]_{\tau_{\rm e}}\,.
} 
Here, we have integrated only over the range of momenta specified in Eq.~\eqref{ks}. Note also that we have neglected a term containing $k^2$ compared to $\m^2a_{\rm e}^2$ which is always valid after $\tau_{\rm c}$ for the range given by Eq. \eqref{ks}.

In the following we find an approximate closed form for the energy density at the end of inflation. The matching time $\tau_{\rm c}$ does not depend on the momentum $k$ and assuming it happens during inflation (i.e. $\tau_{\rm i}<\tau_{\rm c}<\tau_{\rm e}$), we can use the analytic expression \eqref{mode2} for the late time. In the expression \eqref{mode2}, the only $k$ dependence comes from the coefficients $c_\ell$ ($\ell = 1,2$) given in Eq. \eqref{c12}
through the variable $x_{\rm c}=-k\tau_{\rm c}$. From the range of momenta given in Eq.~\eqref{ks}, $x_{\rm c}$ which is the argument of $H^{(1)}_\nu$ in $c_\ell$, can take values between
\eq{
\sqrt{p(p-1)}e^{-(\mathcal{N}-\Delta\mathcal{N})}<x_{\rm c}<\sqrt{p(p-1)}\,.
} 
Further, the argument of the Hankel functions  $H^{(\ell)}_\mu$ in $c_\ell$  is $y_{\rm c}$ which is always smaller than unity. As a result, for the purpose of analytic studies it is reasonable to use the small argument approximation for all the Hankel functions as follows
\eq{
H^{(1)}_\alpha(x)\simeq-i\frac{\Gamma(\alpha)}{\pi}\left(\frac{x}{2}\right)^{-\alpha}\,,\qquad H_\alpha^{(1)}{}'(x)\simeq i\frac{\Gamma(\alpha+1)}{2\pi}\left(\frac{x}{2}\right)^{-\alpha-1}\,,
} 
with $\alpha>0$. For  $H^{(2)}_\alpha$ similar expressions can be obtained noting that $H^{(2)}_\alpha = {H^{(1)}_\alpha}^*$.  

Using the above approximations and after some lines of algebra we find the following approximate expressions for the coefficients \eqref{c12}
\eq{
c_1\simeq c_2\simeq -i\frac{\Gamma(\nu)\Gamma(\mu)}{4\pi}\left(\frac{x_{\rm c}}{2}\right)^{-\nu}\left(\frac{y_{\rm c}}{2}\right)^{-\mu}\,.
}

Plugging these values of $c_\ell$ in the final mode function in  Eq. (\ref{energy-T}) for
$\rho_T(\tau_{\rm e})$ and using  the definition Eq.~\eqref{rhoC}  we obtain 
\eq{
	\label{CT}
	\mathcal{C}_T\simeq\frac{e^{-2\nu\Delta\mathcal{N}}}{2\pi^2}\Big(\frac{\Gamma(\nu)\Gamma(\mu)}{2^{2-\nu-\mu}\pi}\Big)^2\Big(\frac{p}{p-1}\Big)^{\frac{1}{2p}}\int_{x_{\rm min}}^{x_{\rm max}}\dd{x}(L_{\rm e}+R\,K_{\rm e})x^{2-2\nu}\,,
}
where as before, the index ${\rm e}$ means the function is evaluated at $\tau_{\rm e}$. Here, we have defined $L(\tau)$ and $K(\tau)$ which are functions of time and not momentum $k$ as follows
\begin{eqnarray}
\label{LK}
L(\tau) \equiv \pi \big[ p y J_{\mu-1}(y) + (p-1) J_{\mu}(y) \big]^2 \,,\hspace{1cm}
K(\tau) \equiv \pi J_{\mu}(y)^2 \,,
\end{eqnarray}
where $J_{\mu}$ is the Bessel function of the first kind, and the integration domain is given by
\eq{
	x_{\rm min}=\sqrt{p(p-1)}e^{-\mathcal{N}}\,,\qquad x_{\rm max}=\sqrt{p(p-1)}e^{-\Delta\mathcal{N}}\,.
}

For the arguments of the Bessel functions in Eq.~\eqref{LK} at the end of inflation, we have $y_{\rm e}=\sqrt{R}/p>1$. Note that we are mostly interested in the regime where $y_{\rm e}\gg1$ (i.e. $R\gg p^2$) so we can use the large argument approximation of the Bessel functions to write
\eq{
L_{\rm e}\simeq p\sqrt{R}\Big[ 1-\sin \big(\frac{4\sqrt{R}-\pi}{2p}\big)\Big]\,,\qquad K_{\rm e}\simeq \frac{p}{\sqrt{R}}\Big[1+\sin \big(\frac{4\sqrt{R}-\pi}{2p}\big) \Big]\,,
}
which yields $L_{\rm e}+R\,K_{\rm e}\simeq2p\sqrt{R}$. 

As a result, from Eq.~\eqref{CT} we obtain
\eq{
\label{ct}
\mathcal{C}_T\simeq\frac{p[p(p-1)]^{5/2-p}}{\pi^4(4-2p)}\left(\frac{\Gamma(\nu)\Gamma(\mu)}{2^{2-\nu-\mu}}\right)^2\left(\frac{p}{p-1}\right)^{\frac{1}{2p}}
\begin{cases}
-e^{(2p-4)\mathcal{N}}e^{-(p-1)\Delta\mathcal{N}}\,,\hspace{.5cm} p>2\\
e^{-(3-p)\Delta\mathcal{N}}\,,\hspace{1.45cm} 1<p<2
\end{cases} \,,
}
where we have used Eq. \eqref{DeltaN} to express the factor of $\sqrt{R}$ in terms of $\Delta\mathcal{N}$. This is our final expression for the energy density of the transverse modes of the gauge field at the end of inflation.

\subsection{Longitudinal mode}

Now, we compute the same quantity for the longitudinal mode. The equation of motion for the longitudinal mode is given by Eq.~\eqref{MF-EqL} which we repeat as
\begin{equation}\label{MF-L}
	u_{k}'' + \Big( k^2 - \frac{\h''}{\h} + \frac{\m^2a^2}{\f^2} \Big) u_{k} =0 \,,\qquad\quad h = \frac{\m a k}{\sqrt{k^2+\m^2 a^2/\f^2}}\,.
\end{equation}
Here, we have written $\h$, already defined in \eqref{U}, in momentum space. The most straightforward way to find a solution for \eqref{MF-L} is to study its solutions in the two regimes $k^2 \gg \frac{\m^2 a^2}{\f^2}$ and $k^2 \ll \frac{\m^2 a^2}{\f^2}$ separately. 

For short modes $k^2 \gg \frac{\m^2 a^2}{\f^2}$, we have $h\approx \m a$ which leads to $\h''/\h \approx 2/\tau^2$ and Eq.~\eqref{MF-L} simplifies to
\begin{equation}\label{MF-L-short}
	u_{k}'' + \Big( k^2 - \frac{2}{\tau^2} \Big) u_{k} = 0 \,.
\end{equation}
Starting with a Bunch-Davies vacuum the  solution is given by 
\begin{equation}\label{BD-sol}
	u_{k} = \frac{\sqrt{-\pi\tau}}{2} e^{i(1+2{\tilde \nu})\pi/4} H^{(1)}_{{\tilde \nu}}(x)\,,
\end{equation}
where ${\tilde \nu} \equiv 3/2$ and as before the argument of the Hankel function is $x=-k\tau$. 

For long modes $k^2 \ll \frac{\m^2 a^2}{\f^2}$, we have $h\approx k \f$ which gives $\h''/\h = \f''/\f \approx p(p-1)/\tau^2$ and Eq. \eqref{MF-L} simplifies to
\begin{equation}\label{MF-L-long}
	u_{k}'' + \Big( - \frac{p(p-1)}{\tau^2} + \frac{R}{\tau^2} \Big(\frac{\tau_{\rm e}}{\tau}\Big)^{2p} \Big) u_{k} =0 \,,
\end{equation}
which has the following solution
\begin{equation}\label{late-sol}
	u_{k} = \frac{\sqrt{-\pi\tau}}{2} e^{i(1+2{\tilde \nu})\pi/4} \Big[
	{\tilde c}_2 H^{(1)}_{{\tilde \mu}}(y) + {\tilde c}_1 H^{(2)}_{{\tilde \mu}}(y) \Big]; 
	\hspace{1cm} {\tilde \mu} \equiv 1-\frac{1}{2p} \,.
\end{equation}
Note that the above expression is an exact solution of Eq.~\eqref{MF-L-long} which, apart from the  $k^2$ term, is the same as Eq. \eqref{v-eq1} for the  transverse modes. As a result, similar to the transverse case, the second term in the big bracket  of \eqref{MF-L-long} dominates over the first term around and after the time $\tau_{\rm c}$ defined in \eqref{etac}. Such behaviour is encoded in the analytic solution \eqref{late-sol}.  

The two solutions \eqref{BD-sol} and \eqref{late-sol} should be matched at the  time $\tilde{\tau}_{\rm c}$ defined by 
\begin{equation}\label{TN}
k^2=\frac{\m^2 a^2}{\f^2}\quad	 \implies \quad  \frac{\tilde{\tau}_{\rm c}}{\tau_{\rm e}}=\Big(\frac{\sqrt{R}}{-k\tau_{\rm e}}\Big)^{\frac{1}{p+1}}\,.
\end{equation} 
As it is evident from the above definition, unlike $\tau_{\rm c}$, the time $\tilde{\tau}_{\rm c}$ depends on the momentum. To go further, we identify two types of scales. First, short modes, namely $-k\tau_{\rm e}>\sqrt{R}$, for which the critical time $\tilde{\tau}_{\rm c}$ happens after inflation. For these short scales the form of the mode function is well approximated by \eqref{BD-sol} during inflation. Note that we are mostly interested in $R>1$. As a result, these scales never exit the horizon and from Eq.~\eqref{MF-L-short} they are not excited during inflation. Thus, we can safely ignore the contributions  of these scales in the energy density of the longitudinal mode. Second, we have long modes $-k\tau_{\rm e}<\sqrt{R}$ for which the time $\tilde{\tau}_{\rm c}$ occurs during inflation. For these scales the late time behaviour is described by Eq.~\eqref{late-sol}. The unknown coefficients ${\tilde c}_\ell$ are obtained by requiring the continuity of the mode function and its time derivative at $\tilde{\tau}_{\rm c}$. Performing these matching conditions,  we obtain 
\ba
\label{c12tilde}
\tilde{c}_{\ell} = \frac{\mp i \pi}{4 p} \left[  p \tilde{y}_{\rm c} 
H_{\tilde \nu}^{(1)}(\tilde{x}_{\rm c}) \,
H_{\tilde \mu}^{(\ell)}{}'(\tilde{y}_{\rm c}) + \tilde{x}_{\rm c} \,  
H_{\tilde \nu}^{(1)}{}'(\tilde{x}_{\rm c}) \,  H_{\tilde \mu}^{(\ell)}(\tilde{y}_{\rm c}) \right] \, ,
\qquad \ell=1,2
\ea
where a prime denotes derivative with respect to the corresponding argument and by definition we have 
\eq{
	\tilde{x}_{\rm c}\equiv -k\tilde{\tau}_{\rm c}\,,\qquad \tilde{y}_{\rm c}\equiv \frac{\sqrt{R}}{p}\left(\frac{\tau_{\rm e}}{\tilde{\tau}_{\rm c}}\right)^{p}=\frac{\tilde{x}_{\rm c}}{p}\,.
}
One important difference compared to the transverse modes is that $\tilde{y}_{\rm c}$ now depends on the momentum $k$ while the corresponding quantity $y_{\rm c}$ for the transverse modes is independent of $k$. 

Here we should comment that the analytic mode function obtained by matching early and late time solution at $\tilde{\tau}_{\rm c}$ is not entirely an accurate expression for the mode function. The reason is that for a time interval around $\tilde{\tau}_{\rm c}$, the equation of motion is not in the form given by neither \eqref{MF-L-short} nor \eqref{MF-L-long} but takes a more complicated form. We have checked numerically that although the qualitative behaviour of our analytic result is accurate, this causes an $\order{1}$ discrepancy in the amplitude of the mode function at the end of inflation for the range of parameters of our interest. It is possible to improve the analytic result by solving the equation for the mode function around the time $\tilde{\tau}_{\rm c}$ and then match the result with both early and late solutions given above. However, the resulting expression would be much more complicated which is not worth bothering the reader with. As a result, we continue exploiting the above analytic expression understanding that for a high precision result one needs to perform numerical analysis.    
 
We are only interested in the range of momenta that experience growth through breaking the adiabaticity condition which are given by (see appendix \ref{app} for the details)
\begin{eqnarray}\label{limits-k2}
x_{\min} = \sqrt{2}\, e^{-{\cal N}}\,, \hspace{1cm}
x_{\max} = \begin{cases}
\sqrt{p(p-1)}e^{-\Delta\mathcal{N}} \hspace{2.75cm} p>2
\\ 
2^{\frac{p+1}{2p}}[p(p-1)]^{-\frac{1}{2p}}e^{-\Delta\mathcal{N}} \hspace{.85cm} 1<p<2
\end{cases} \,,
\end{eqnarray}
where $x_{\min} = -k_{\min}\tau_{\rm e}$ and $x_{\max} = -k_{\max}\tau_{\rm e}$ as before and the upper bound corresponds to the momentum for which $-k_{\max}\tilde{\tau}_{\rm c}(k_{\max})=\sqrt{2}$. Correspondingly, the energy density of the longitudinal mode at the end of inflation \eqref{rhoT4} is approximated by 
\eq{
	\rho_L(\tau_{\rm e})\approx\frac{1}{2a_{\rm e}^4}\int_{k_{\rm min}}^{k_{\rm max}}\frac{\dd[3]{k}}{(2\pi)^3}
	\Big[ \big|{\big(\frac{u_{k}}{\f}\big)'} \big|^2+\m^2a_{\rm e}^2\abs{u_{k}}^2\Big]_{\tau_{\rm e}}\,,
}  
where we have used the fact $h(\tau_{\rm e})\simeq k f(\tau_{\rm e})=k$ and also  neglected the $k^2$ term in comparison to the mass term for the momenta in the range given in Eq. \eqref{limits-k2}. Note that the mode function at the end of inflation is given by Eq.~\eqref{late-sol} in which the coefficients ${\tilde c}_\ell$ given in Eq. \eqref{c12tilde}. 

To proceed further, we note that from \eqref{limits-k2} we have
\begin{equation}
\begin{cases}
\sqrt{2}\, e^{-(\mathcal{N}-\Delta\mathcal{\tilde N})} < 
\tilde{x}_{\rm c} < \sqrt{p(p-1)}e^{\Delta\mathcal{\tilde N}-\Delta\mathcal{N}} \hspace{2.75cm} p>2
\\ 
\sqrt{2}\, e^{-(\mathcal{N}-\Delta\mathcal{\tilde N})} < 
\tilde{x}_{\rm c} <
2^{\frac{p+1}{2p}}[p(p-1)]^{-\frac{1}{2p}}e^{\Delta\mathcal{\tilde N}-\Delta\mathcal{N}} \hspace{.85cm} 1<p<2
\end{cases} \,,
\end{equation}
where $\Delta{\cal\tilde N}=\ln\frac{{\tilde \tau}_{\rm c}}{\tau_{\rm e}}$. As a result, $\tilde{x}_{\rm c}$ is effectively a small quantity. More precisely, since $\Delta{\cal\tilde N} - \Delta{\cal N} =\ln\frac{{\tilde \tau}_{\rm c}}{\tau_{\rm c}}$, we have $\Delta\mathcal{\tilde N}<\Delta\mathcal{N}$ and $\Delta\mathcal{\tilde N}>\Delta\mathcal{N}$ for ${\tilde \tau}_{\rm c}< \tau_{\rm c}$ and ${\tilde \tau}_{\rm c} > \tau_{\rm c}$ respectively. Therefore, in the case of ${\tilde \tau}_{\rm c}< \tau_{\rm c}$ the approximation $\tilde{x}_{\rm c}\ll1$ is more accurate than the case ${\tilde \tau}_{\rm c} > \tau_{\rm c}$. So, similarly to the case of the transverse modes, we use the small argument approximation of the Hankel functions yielding 
\eq{
	{\tilde c}_1\simeq {\tilde c}_2\simeq -i\Big(\frac{p+1}{p}\Big)\frac{\Gamma({\tilde \nu})\Gamma({\tilde \mu})}{4\pi}\left(\frac{{\tilde x}_{\rm c}}{2}\right)^{-{\tilde \nu}}\left(\frac{{\tilde y}_{\rm c}}{2}\right)^{-{\tilde \mu}}\,.
}

Performing  a similar analysis which led to Eq.~\eqref{ct}  we obtain 
\eq{
\label{cl}
\mathcal{C}_L\simeq\frac{p(p+1)^3}{\pi^4(4-2p)}
\Big(\frac{\Gamma(\tilde{\nu})\Gamma(\tilde{\mu})}{2^{2-\tilde{\nu}-\tilde{\mu}}}\Big)^2
\frac{\big( p(p-1)\big)^{\frac{1}{2}-\frac{3}{2p}}}{p^{\frac{1}{p}}2^{-\frac{2p+5}{1+p}}}
\begin{cases}
- [p(p-1)]^{\frac{2-p}{p(p+1)}} e^{\frac{2p-4}{p+1}\mathcal{N}}
e^{\frac{p^2-4p+1}{p+1}\Delta\mathcal{N}}\hspace{.6cm} p>2\\ 
2^{\frac{2-p}{p(p+1)}}  e^{-(3-p)\Delta{\cal N}}\hspace{3cm} 1<p<2
\end{cases},
}
where $\mathcal{C}_L$ is defined in Eq. \eqref{rhoC} as $\rho_L(\tau_{\rm e})=\mathcal{C}_LH_{\rm e}^4$.

\section{Dark matter relic density}
\label{relic-sec}

In the previous section we obtained the energy density of the gauge field for both transverse and longitudinal modes at the end of inflation. In this section we use the above results to compute its present energy density after getting redshifted after inflation due to cosmic expansion.

First of all, note that the energy of the produced dark photons during inflation must be sub-dominant compared to the inflationary background. Besides, the backreaction of the gauge field on the inflaton slow-roll  
dynamics must be negligible. As  discussed in Refs. \cite{Nakai:2020cfw,Salehian:2020asa}, in our phenomenological model with the coupling \eqref{f1}, the backreaction condition on the inflaton field's slow-roll  dynamics 
is stronger. This condition constrains the value of $p$. For instance, similar to \cite{Salehian:2020asa},  for $\mathcal{N}=50$ e-folds of inflation  one must have $p\lesssim2.2$  in order to avoid any backreaction. 

In addition, there are strong constraints from the dark matter isocurvature perturbations. In this setup the origin of dark matter is different than the origin of baryonic matter and radiation which are generated from the decay of inflaton at the end of inflation. Therefore, dark matter isocurvature will inevitably be generated. This constraint was studied in \cite{Salehian:2020asa} yielding to the bound $p< 2.01$ in 
our setup with the coupling \eqref{f1}. Interestingly, we see that the constraint on $p$
from the isocurvature perturbation is stronger than the backreaction condition. 

Assuming that the above conditions are satisfied, we must compute the present time energy density of the gauge field. Note that since we assume a large value for the parameter $R$, the dark photons become non-relativistic right after inflation and their energy density decrease like $a^{-3}$ in the subsequent cosmic history. The present energy density of dark photon can be approximated as \cite{Salehian:2020asa}  
\eq{
\label{omg}
\Omega_{A,0}=0.5\times10^{-4}\left(\frac{T_{\rm r}}{10^{12}{\rm GeV}}\right)^5\mathcal{C}(p,{\cal N}, \Delta{\cal N})\,,
}
where $\mathcal{C}=\mathcal{C}_T+\mathcal{C}_L$ in which $\mathcal{C}_T$ and $\mathcal{C}_L$ are given by Eqs.~\eqref{ct} and \eqref{cl} respectively and $T_{\rm r}$ is the reheating temperature. In deriving above result we have assumed an instantaneous reheating for simplicity.

\begin{figure}[t!]
	\centering
	\includegraphics[width=0.45\textwidth]{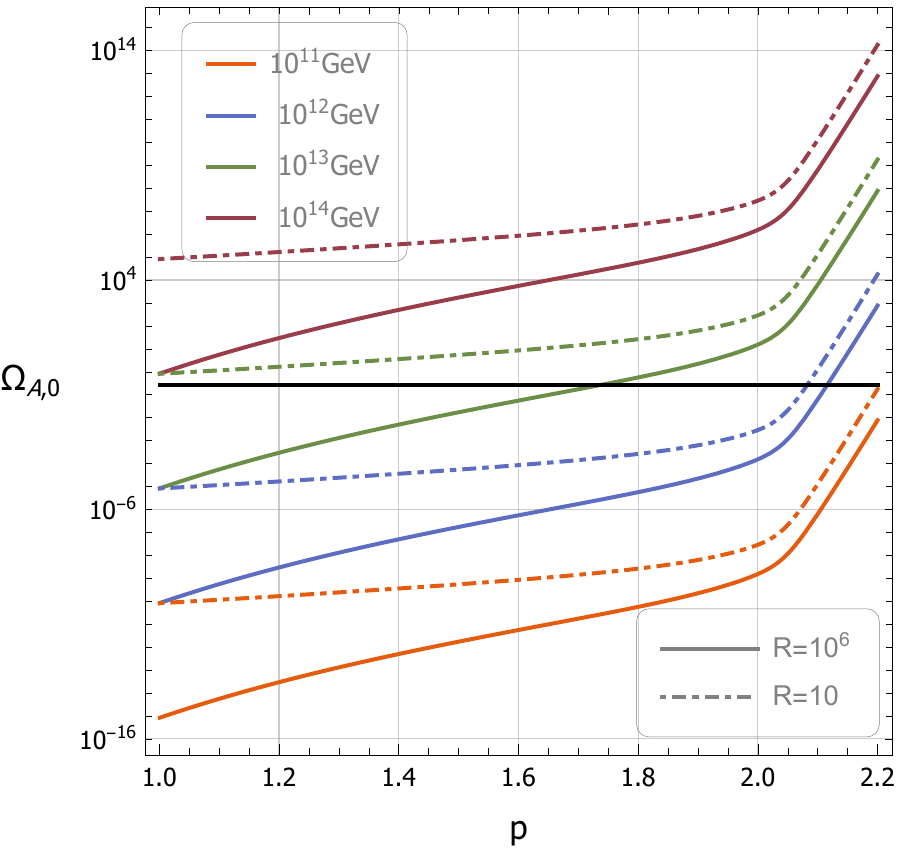}
	\hspace{8mm}
	\includegraphics[width=0.445\textwidth]{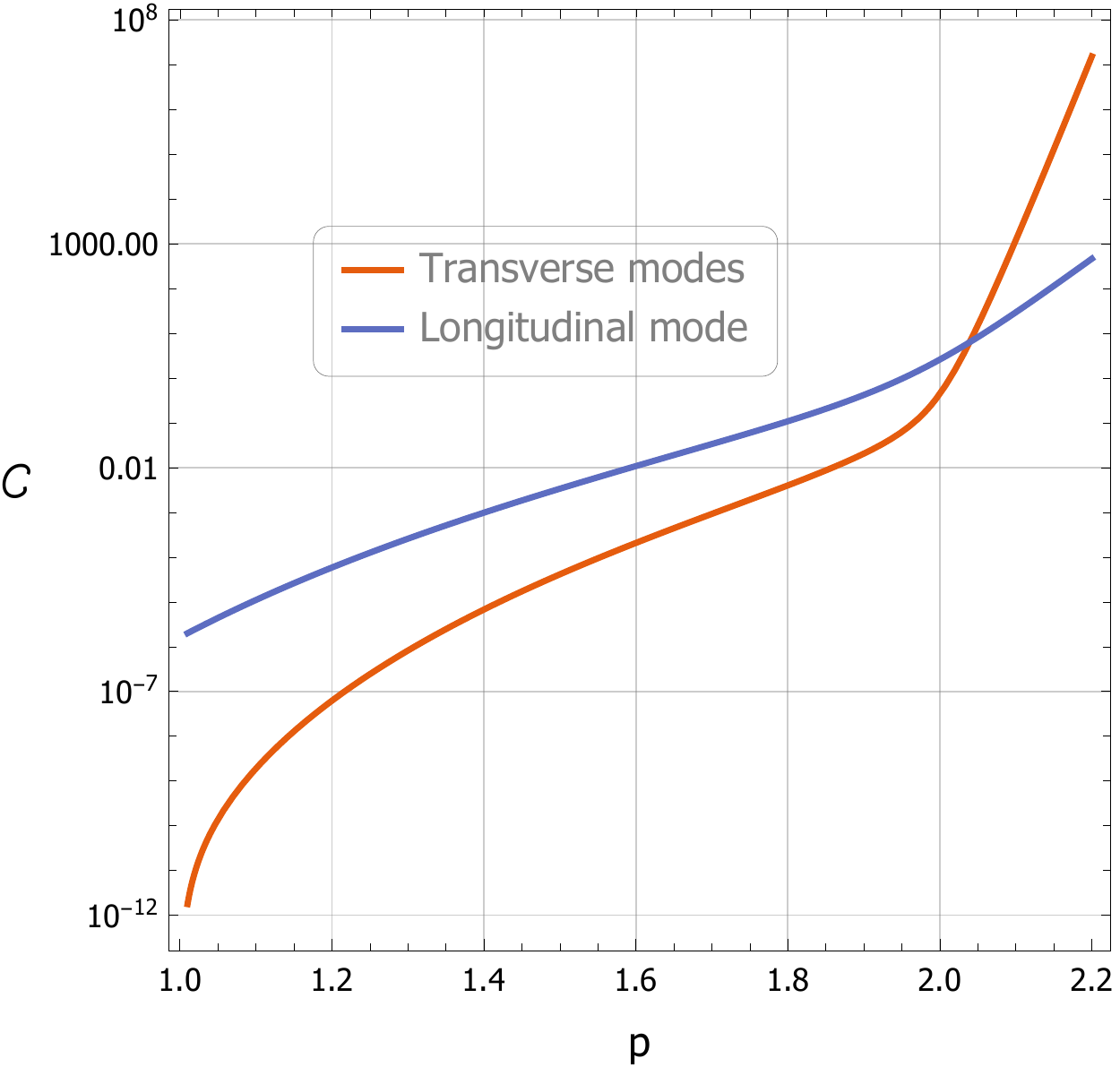}
	\caption{Left: present energy density of dark photons given by Eq.~\eqref{omg} as a function of $p$ for $R=10$ (dashed-dotted curves) and $R=10^6$ (solid curves) for different reheat temperatures with $\mathcal{N}=50$. The solid black horizontal line represents the observed dark matter density $\Omega_D=0.27$. Right: comparison of the  contributions of the  transverse and longitudinal modes to the energy density for $R=10^6$. }
	\label{fig:relic}
\end{figure}

We are mostly interested in the case $R>1$ where the dark matter is non-relativistic right after inflation. The left panel of Fig.~\ref{fig:relic} shows the relic energy density of dark photons for two values of $R=10$ (dashed-dotted curves) and $R=10^6$ (solid curves) for different reheat temperatures. We see that the dependence of the relic energy density on $R$ is stronger for $p<2$ than $p>2$. For a fixed value of $p$, the reheat temperature is higher for larger values of $R$ in order for the dark photons furnish the observed dark matter relic density.  If the dark photon comprises the whole dark matter energy density, the reheat temperature is typically in the range  $10^{12} \mathrm{GeV} <  T_{\mathrm r} < 10^{14} \mathrm{GeV}$.

In the right panel of Fig.~\ref{fig:relic} we have compared the contribution of the longitudinal mode in the total energy density to that of the transverse modes. 
As we see the contribution from the longitudinal mode for $p>2$ is negligible in agreement with the results of \cite{Nakai:2020cfw,Salehian:2020asa}. However, for $1<p<2$, we see that the longitudinal mode is dominant. This is a new result and roughly in agreement with the results of Ref. \cite{Graham:2015rva} where it is shown that for $p=0$, the observed dark matter relic can be achieved from the longitudinal mode. But, let us look more carefully to the origin of these results. 
Note that one main difference between the longitudinal mode and transverse modes is that the particle production for the longitudinal mode is governed initially by the term $-2/\tau^2$ and then the term $-p(p-1)/\tau^2$ while for the transverse modes this is always the term $-p(p-1)/\tau^2$ which is responsible for growth. For $p>2$, the factor $-2/\tau^2$ has less weight compared to the factor $-p(p-1)/\tau^2$ and thus the longitudinal mode experiences less growth compared to the transverse modes. However for $1<p<2$, the stronger growth happens for the longitudinal mode.

Throughout this paper, we always assumed that $R>1$ and we did not study the case $R<1$ as it is already studied in the literature  \cite{Nakayama:2019rhg,Nakayama:2020rka,Nakai:2020cfw,Salehian:2020asa}. However, some comments for the case $R<1$ are in order. For $p>2$ and $R<1$, it is shown that one can always neglect the longitudinal mode \cite{Nakai:2020cfw} while this scenario is restricted to $p<2.01$ from the bound on the isocurvature perturbations \cite{Salehian:2020asa}. In the case of $1<p<2$ and $R<1$ the short scale modes with $-k\tau_{\rm e}>\sqrt{R}$ give the main contribution to the energy density of the longitudinal mode and it can be shown that this contribution is also larger than the contribution coming from the transverse modes. Therefore, we can have a scenario of light vector dark matter in which the whole contribution comes from the longitudinal mode. This scenario is qualitatively similar to the model considered in Ref. \cite{Graham:2015rva} which corresponds to the special case of $p=0$.

\section{Summary and discussion}\label{summary}

In this paper we considered a new mechanism of vector dark matter production during inflation. The setup contains a complex inflaton field which is charged under the dark $U(1)$ gauge field. As the inflaton field rolls towards the global minimum of the potential the gauge field acquires mass due to the Higgs mechanism. In order for the gauge field to play the roles of dark matter we require the potential to have a global minimum with a non-zero vacuum expectation value of the field. The accumulated energy density associated with the excited quantum fluctuations of the transverse and the longitudinal modes play the roles of the observed dark matter energy density.  We  mostly concentrated  in the cases where the vector field is massive during inflation with $R>1$ so the dark photons become non-relativistic right after inflation. This is motivated from simple models of inflation such as the Higgs symmetry breaking potential which is under the category of large field inflation, i.e. fields value larger than $M_{\rm {Pl}}$. However, there is no fundamental reason for the potential be in the category of large field setup, so one may consider potentials in which the field displacement is not large so a smaller value of $R$ is also possible.

In order to get energy from inflaton and prevent the decay of the gauge field energy density, we have added a coupling between the inflaton field and the gauge field through the gauge kinetic function. Such a coupling not only pumps energy from the inflaton sector to the gauge field sector but also reduces the effective mass of the vector field so the quanta of the vector field can be excited efficiently during  most of the period of inflation. An interesting feature of this setup is that the vector field is massive during the entire period of inflation so the longitudinal mode also contributes to the dark matter relic density. For a gauge kinetic function with a power law time dependence with the free parameter $p$ we have calculated the relic energy density associated with the transverse and the longitudinal modes. Requiring a negligible backreaction on the background inflaton dynamics and imposing the dark matter isocurvature constraint implies that $p< 2.01$. We have shown that the observed dark matter relic density can be generated with the reheat temperature in the range $10^{12} \mathrm{GeV} <  T_{\mathrm r} < 10^{14} \mathrm{GeV}$. For a fixed value of $p$ models with larger values of $R$ require higher reheat temperature to obtain the observed dark matter relic density.

The current analysis can be extended in various directions. First, one can consider the setup with $R<1$ so the vector field is not non-relativistic at the end of inflation. In this case we have to wait until the time $t_{\mathrm{NR}}$ when the mode function becomes non-relativistic and from then the energy density of the dark photon can be carried to the time of matter and radiation equality as the seed of dark matter. This case was studied in details in   \cite{Salehian:2020asa} and a similar analysis can be performed here as well. The second direction is to look for the dark photon productions during (p)reheating. More specifically, the oscillations of the inflaton field near its minimum at the end of inflation induce an oscillatory coupling to the gauge field via the gauge kinetic function $f(\varphi)$. This can excite the quanta of dark photon via parametric resonance. This channel of particle production is more significant for the case of $R<1$ where the dark photon is light while such particle creation is expected to be negligible for the case $R>1$ as the excitations of heavy dark photon may not be efficient at the end of inflation. This is an interesting question which is beyond the scope of this work. Finally, while we have considered the inflaton as a complex scalar field whose vev gives mass to the dark photon, the analysis in the present paper can be applied to more general setups in which $\phi$ is not necessarily the inflaton as far as the $U(1)$ gauge symmetry is broken for the whole period of inflation. Therefore it is certainly intriguing to explore other possibilities.

\vspace{0.7cm}

{\bf Acknowledgments:} 
The work of M.A.G. was supported by Japan Society for the Promotion of Science Grants-in-Aid for international research fellow No. 19F19313. The work of S.M. was supported in part by Japan Society for the Promotion of Science Grants-in-Aid for Scientific Research No.~17H02890, No.~17H06359, and by World Premier International Research Center Initiative, MEXT, Japan. 

\vspace{0.7cm}

\appendix
\section{Growing momenta for longitudinal mode} \label{app}

As we discussed in the main text, the short modes with $-k\tau_{\rm e}>\sqrt{R}$ do not grow and we do not consider them further here. For the modes with $-k\tau_{\rm e}<\sqrt{R}$ the critical time $\tilde{\tau}_{\rm c}$ occurs during inflation. We define the number of e-folds from $\tilde{\tau}_{\rm c}$ until the time of end of inflation from \eqref{TN} as
\ba
\Delta\tilde{\mathcal{N}}\equiv \ln\frac{{\tilde \tau}_{\rm c}}{\tau_{\rm e}} 
=\frac{1}{p+1}\ln(\frac{\sqrt{R}}{-k\tau_{\rm e}})\,.
\ea 
Depending on the parameters, $\tilde{\tau}_{\rm c}$ can occur before or after $\tau_{\rm c}$. To see this, we look at
\eq{
\Delta\mathcal{N}-\Delta\tilde{\mathcal{N}} = \ln\frac{\tau_{\rm c}}{{\tilde \tau}_{\rm c}} 
=\frac{1}{p+1}\ln(\frac{-k\tau_{\rm e}}{\sqrt{p(p-1)}e^{-\Delta\mathcal{N}}})\,.
}  

First we consider $-k\tau_{\rm e}>\sqrt{p(p-1)}e^{-\Delta\mathcal{N}}$ for which $\tilde{\tau}_{\rm c}$ happens after $\tau_{\rm c}$. As a result, for these scales tachyonic instability is possible only if the expression in the bracket in (\ref{MF-L-short}) becomes negative before ${\tilde \tau}_{\rm c}$. Namely, the tachyonic instability in this case requires $-k\tilde{\tau}_{\rm c}<\sqrt{2}$, which is rewritten as $-k\tau_{\rm e}< 2^{\frac{p+1}{2p}}[p(p-1)]^{-\frac{1}{2p}}e^{-\Delta\mathcal{N}}$ and thus is inconsistent with the our primary condition $-k\tau_{\rm e}>\sqrt{p(p-1)}e^{-\Delta\mathcal{N}}$ for $p>2$. Therefore there is no chance to have any growing mode for $-k\tau_{\rm e}>\sqrt{p(p-1)}e^{-\Delta\mathcal{N}}$ if $p>2$. For $p<2$, however, we have growing modes for
\begin{eqnarray}\label{limit1}
\sqrt{p(p-1)}e^{-\Delta\mathcal{N}}<-k\tau_{\rm e} < 2^{\frac{p+1}{2p}}[p(p-1)]^{-\frac{1}{2p}}e^{-\Delta\mathcal{N}} \,,
\end{eqnarray}

Second, we consider the range $-k\tau_{\rm e}<\sqrt{p(p-1)}e^{-\Delta\mathcal{N}}$ for which $\tilde{\tau}_{\rm c}$ happens before $\tau_{\rm c}$. For these scales the equation of the mode function can be approximated as
\eq{ \label{eqn:taui<tau<tauctilde}
u_{k}'' + \Big( k^2 - \frac{2}{\tau^2} \Big) u_{k} =0 \,,\qquad \tau_{\rm i}<\tau<\tilde{\tau}_{\rm c}
}
\eq{
	u_{k}'' + \Big(- \frac{p(p-1)}{\tau^2} \Big) u_{k} =0 \,,\qquad \tilde{\tau}_{\rm c}<\tau<\tau_{\rm c}
}
\eq{
	u_{k}'' + \Big(\frac{R}{\tau^2}\left(\frac{\tau_{\rm e}}{\tau}\right)^{2p} \Big) u_{k} =0 \,,\qquad \tau_{\rm c}<\tau<\tau_{\rm e}\,.
}
We thus find the following range of momenta for the growing modes
\ba\label{limit2}
\sqrt{2}\, e^{-{\cal N}}< - k\tau_{\rm e} < \sqrt{p(p-1)}e^{-\Delta\mathcal{N}} \,.
\ea

From Eqs. \eqref{limit1} and \eqref{limit2} we find the range of momenta for the all growing modes as follows
\begin{equation}\label{range-long}
x_{\min} = \sqrt{2}\, e^{-{\cal N}}\,, \hspace{1cm}
x_{\max} = \begin{cases}
\sqrt{p(p-1)}e^{-\Delta\mathcal{N}} \hspace{2cm} p>2
\\ 
2^{\frac{p+1}{2p}}[p(p-1)]^{-\frac{1}{2p}}e^{-\Delta\mathcal{N}} \hspace{.85cm} p<2
\end{cases} \,,
\end{equation} 
which we use to find the energy density of the produced longitudinal dark photons. Here, the lower bound corresponds to the mode for which the expression in the bracket in (\ref{eqn:taui<tau<tauctilde}) changes its sign at the beginning of inflation.

\bibliography{ref} 
\bibliographystyle{JHEP}

\end{document}